\documentclass[a4paper,11pt]{article}
\usepackage{jheppub}
\usepackage{lineno}
\usepackage{tikz}
\usepackage{cleveref}
\usepackage{xspace}
\usepackage{caption}
\usepackage{subcaption}
\usepackage{glossaries}
\glsdisablehyper
\usepackage{siunitx}
\usepackage{hyperref}
\usepackage{listings}

\crefname{lstlisting}{code}{code}
\Crefname{lstlisting}{Code}{Code}

\lstdefinestyle{pythoncode}{
  language=Python,
  basicstyle=\fontsize{9}{10}\selectfont\ttfamily
}

\usepackage[sorting=none,isbn=false,eprint=false]{biblatex}
\AtEveryBibitem{%
  \clearfield{date}%
  \ifentrytype{software}{%
  }{%
    \clearfield{url}%
    \clearfield{urldate}%
    \clearfield{urlyear}%
    \clearfield{urlmonth}%
    \clearfield{urlday}%
  }%
}
\newcommand{\FlexCAST}{\textsc{FlexCAST}\xspace}
\newcommand{\RECAST}{\textsc{RECAST}\xspace}
\newcommand{\REANA}{\textsc{REANA}\xspace}
\newcommand{\MadAnalysis}{\textsc{MadAnalysis}\xspace}
\newcommand{\luigi}{Luigi\xspace}
\newcommand{\law}{Law\xspace}
\newcommand{\task}[1]{\texttt{#1}\xspace}
\newcommand{\dataset}[1]{\texttt{#1}\xspace}
\newcommand{\mjj}{\ensuremath{m_{JJ}}\xspace}

\newacronym{SM}{SM}{Standard Model of particle physics}
\newacronym{BSM}{BSM}{beyond the Standard Model}
\newacronym{AD}{AD}{Anomaly Detection}
\newacronym{DAG}{DAG}{Directed Acyclic Graph}
\newacronym{ML}{ML}{Machine Learning}
\newacronym{MLOps}{MLOps}{Machine Learning Operations}
\newacronym{HEP}{HEP}{high-energy physics}
\newacronym{SR}{SR}{signal region}
\newacronym{SB}{SB}{sideband region}
\newacronym{CWOLA}{\textsc{CWoLa}}{Classification Without Labels}
\newacronym{law}{\law}{Luigi Analysis Workflow}

\title{\FlexCAST: Enabling Flexible Scientific Data Analyses}

\author[a,b]{Benjamin Nachman}
\emailAdd{nachman@stanford.edu}
\author[a,c]{and Dennis Noll}
\emailAdd{dnoll@lbl.gov}
\affiliation[a]{Fundamental Physics Directorate, SLAC National Accelerator Laboratory,\\Menlo Park, CA 94025, USA}
\affiliation[b]{Department of Particle Physics and Astrophysics, Stanford University,\\Stanford, CA 94305, USA}
\affiliation[c]{Physics Division, Lawrence Berkeley National Laboratory,\\Berkeley, CA 94720, USA}

\abstract{The development of scientific data analyses is a resource-intensive process that often yields results with untapped potential for reuse and reinterpretation. In many cases, a developed analysis can be used to measure more than it was designed for, by changing its input data or parametrization. Existing reinterpretation frameworks, such as \RECAST, enable analysis reinterpretation by preserving the analysis implementation to allow for changes of particular parts of the input data. We introduce \FlexCAST, which generalizes this concept by preserving the analysis design itself, supporting changes to the entire input data and analysis parametrization. \FlexCAST is based on three core principles: modularity, validity, and robustness. Modularity enables a change of the input data and parametrization, while validity ensures that the obtained results remain meaningful, and robustness ensures that as many configurations as possible yield meaningful results. While not being limited to data-driven machine learning techniques, \FlexCAST is particularly valuable for the reinterpretation of analyses in this context, where changes in the input data can significantly impact the parametrization of the analysis. Using a state-of-the-art anomaly detection analysis on LHC-like data, we showcase \FlexCAST's core principles and demonstrate how it can expand the reach of scientific data analysis through flexible reuse and reinterpretation.}

\begin{document}
\maketitle
\flushbottom

\section{Introduction and Motivation}
\label{sec:intro}

Scientific data analyses typically require significant resources, expertise, and time to develop, validate, and execute.
Despite this substantial effort, they are often used only once, specifically for the measurement or hypothesis for which they were initially designed.
However, analyses frequently contain valuable methods, workflows, and insights that could be reused or adapted to answer the same or related scientific questions on new data or to reinterpret existing data in the light of new questions.
For example, in \gls{HEP}, an analysis originally designed to search for a particular signal model could potentially be reused to test for the existence of other signal models not considered during the design of the analysis.
To facilitate such reuse and reinterpretation, specialized frameworks have emerged, providing strategies and tools that allow analyses to be preserved and adapted effectively.
One prominent example of such a framework is \RECAST~\cite{cranmer_recast_2011,cranmer_yadage_2017,cranmer_analysis_2018}, which enables physicists to preserve analyses in a reusable form to allow reinterpretation.
The framework has been used to preserve many analyses yielding new scientific results~\cite{atlas_collaboration_recast_2019,atlas_collaboration_reinterpretation_2020,collaboration_search_2023,collaboration_atlas_2024,collaboration_statistical_2024,collaboration_combination_2024}.

\RECAST enables the reuse of existing analyses by allowing variations in a specific part of the input data, namely the signal model.
In this approach, the analysis parametrization - the specific methodology of the analysis, including, for example, the event selection criteria or \gls{ML} model weights - remains unchanged between the design of the analysis and the reinterpretation.
Because the analysis parametrization itself remains the same, the validity of the analysis, which was established during the original publication process, does not need to be demonstrated again \footnote{While this assumption is made in the \RECAST approach, it is in practice not necessarily checked explicitly.}.
Consequently, publishing results for a new signal model becomes comparatively straightforward, as most of the required validation checks have already been performed.

For certain analyses, however, it becomes necessary to modify one or both, significant proportions of the input data and the analysis parametrization, during the reinterpretation process to ensure that the analysis remains sensitive and valid.
Consider, for example, an analysis looking for an unknown signal that uses a specific size of a signal window in an invariant mass spectrum.
The optimal size of the signal window should surely be determined based on the specific signal that is used in the reinterpretation process.
Another scenario where both input data and parametrization need adjustment is when the analysis algorithms themselves, or parts of them, depend on the input data.
This is, for example, inherently true for data-driven \gls{ML} methods, where the analysis is not static and changes with different input data.

A prime example of a class of physics analyses where both scenarios are realized is the class of \gls{AD} analyses that rely on training classifiers to distinguish potential signals from background events.
First, they often require customization to the class of signal models they are looking for.
Second, their classifiers must learn to recognize patterns from the analyzed dataset, which correspond to a potential, a priori unknown realization of such a signal.
This dual dependency - on both the analysis parametrization and the input data - highlights the need for a flexible and adaptable reinterpretation framework.

We introduce \FlexCAST, a framework aimed at enabling changes in the entire input data and the complete analysis parametrization.
The primary focus of \FlexCAST is on the analysis steps between detector reconstruction and the final scientific results, such as event selection or classification using \gls{ML} methods.
While at first glance \FlexCAST might appear similar to the original \RECAST approach, it represents a fundamental shift in how analyses are preserved and reinterpreted: whereas \RECAST preserves and reinterprets analyses as they were originally designed, \FlexCAST instead preserves the design of the analysis itself.
The following chapters will describe in detail how this fundamental shift can be realized in practice.
Despite this conceptual difference, \FlexCAST retains the key features of \RECAST, such as static software environments, or the API, which enables collaborative control over the approval of new results.

While we discuss \FlexCAST in the context of \gls{HEP}, the \FlexCAST principles can benefit a wide range of scientific data analyses and the framework is particularly valuable whenever the analysis itself depends significantly on the input data - for example, when repeating a data-driven analysis on a new experimental or synthetic dataset.
And although \FlexCAST is not inherently tied to \gls{ML}, its principles are especially relevant for analyses that employ data-driven \gls{ML} strategies, as these approaches inherently depend on and adapt to the input data.

While reinterpretation tools like \MadAnalysis~\cite{conte_madanalysis_2013} provide great value in the extensive study of phenomenological models, they often only provide limited accuracy in the light of the reinterpretation of real experimental data.
Additionally, these tools are not explicitly tailored to analysis strategies that strongly rely on the input data, like \gls{ML}-driven \gls{AD}.

This paper is structured as follows:
\Cref{sec:design} introduces the formal definitions and core design principles of \FlexCAST and \cref{sec:implementation} describes how these principles can be implemented.
Next, \cref{sec:study} demonstrates the practical utility of the implemented framework through a flexible reinterpretation of a state-of-the-art \gls{AD} analysis on LHC-like data.
Finally, \cref{sec:conclusion} summarizes our findings and provides an outlook on future directions.

\section{Design Principles}
\label{sec:design}

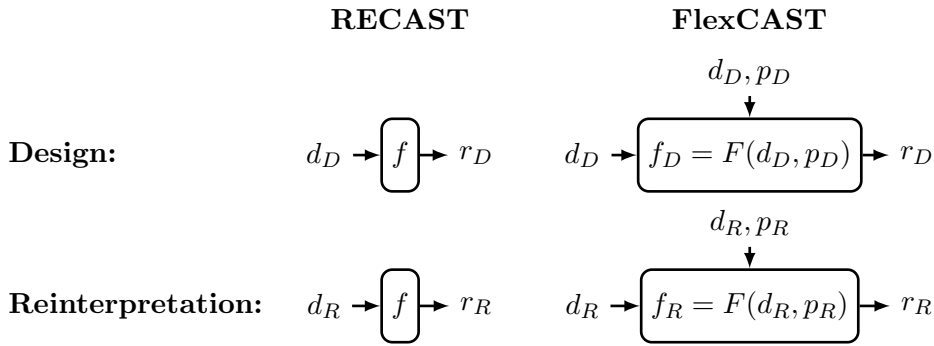
\begin{figure}[htbp]
\centering
\tikzset{
    block/.style={rectangle, draw, text centered,
        rounded corners, minimum height=2.5em, fill=#1},
    block/.default=none,
    line/.style={draw=#1, -Latex},
    line/.default=black,
    every path/.style={line width=1pt}
}

\begin{tikzpicture}[node distance=2cm]
  \node (recast) at (1,1.8) {\textbf{RECAST}};
  \node (flexcast) at (5.6,1.8) {\textbf{FlexCAST}};
  \node[right] (design) at (-4.3,0) {\textbf{Design:}};
  \node[right] (reinterpretation) at (-4.3,-2) {\textbf{Reinterpretation:}};

  \node (D_old) at (0,0) {$d_{D}$};
  \node[block] (f_Recast) at (1,0) [draw, rectangle] {$f$};
  \node (R_old) at (2,0) {$r_{D}$};

  \node (D_new) at (0,-2) {$d_{R}$};
  \node[block] (f_Recast_new) at (1,-2) [draw, rectangle] {$f$};
  \node (R_new) at (2,-2) {$r_{R}$};

  \draw[-latex] (D_old) -- (f_Recast);
  \draw[-latex] (f_Recast) -- (R_old);
  \draw[-latex] (D_new) -- (f_Recast_new);
  \draw[-latex] (f_Recast_new) -- (R_new);

  \node (D_old_2) at (3.4,0) {$d_{D}$};
  \node (p_old) at (5.6, 1.1) {$d_{D}, p_{D}$};
  \node[block] (f_Flexcast_old) at (5.6,0) [draw, rectangle] {$f_{D} = F(d_{D}, p_{D})$};
  \node (R_old_2) at (7.8,0) {$r_{D}$};

  \draw[-latex] (D_old_2) -- (D_old_2 -| f_Flexcast_old.west);
  \draw[-latex] (p_old) -- (f_Flexcast_old.north);
  \draw[-latex] (f_Flexcast_old) -- (R_old_2);

  \node (D_new_2) at (3.4,-2) {$d_{R}$};
  \node (p_new) at (5.6, -0.9) {$d_{R}, p_{R}$};
  \node[block] (f_Flexcast_new) at (5.6,-2) [draw, rectangle] {$f_{R} = F(d_{R}, p_{R})$};
  \node (R_new_2) at (7.8,-2) {$r_{R}$};

  \draw[-latex] (D_new_2) -- (D_new_2 -| f_Flexcast_new.west);
  \draw[-latex] (p_new) -- (f_Flexcast_new.north);
  \draw[-latex] (f_Flexcast_new) -- (R_new_2);

\end{tikzpicture}
\caption{
    This figure illustrates the difference between the two discussed approaches to analysis reinterpretation. In \RECAST, the analysis function ($f$) remains unchanged from the design phase ($D$) to the reinterpretation phase ($R$). In contrast, \FlexCAST allows the analysis to be adapted using data ($d$) and additional parameters ($p$).
    \label{fig:schema}
}
\end{figure}

This section introduces a theoretical framework for flexible analysis preservation and outlines the three core design principles of \FlexCAST.
The theoretical framework of \FlexCAST can be understood through the lens of functions and functionals following \cref{fig:schema}.
In the context of \RECAST, an analysis can be viewed as a carefully defined function, denoted as $f$, which takes input data $d$ and produces a result $r$, as $r = f(d)$.
Here, the data typically refers to the experimentally recorded and simulated data used during the design of the analysis ($d_D$), as well as additionally the simulated data for a new hypothesized signal used during the reinterpretation process ($d_R$).
On the other hand, \FlexCAST introduces the concept of a functional, denoted as $F$, which returns an analysis $f$, dependent on the data $d$ and potentially additional parameters $p$: $f = F(d, p)$.
The analysis $f$ is then a deterministic function that transforms the input data into the final result.
During the reinterpretation process, the data and the parameters change, $d_R$ and $p_R$, and the \FlexCAST functional will return a new function $f_R$ which can be applied on the changed data to produce a new result, $r_R = f_R(d_R)$.
So while \RECAST primarily preserves the analysis function, \FlexCAST preserves the analysis functional.
A user not only implements the analysis strategy but also the logic behind how they arrived at a particular analysis strategy, including the steps taken to optimize it.

To illustrate this concept with an example, consider that a functional $F$ might include complex tasks such as optimizing the width of a window on an invariant mass spectrum or training a neural network~\cite{collins_extending_2019}.
In contrast, the resulting function $f$ would simply deterministically apply this predetermined selection cut or trained network to the input data.
This distinction highlights how \FlexCAST enables a flexible approach to analysis preservation, where the core logic of the analysis design is separated from its specific application.

\FlexCAST employs three core design principles to enable this flexibility in the preservation of scientific data analyses:
\begin{equation*}
  \textbf{Flexibility} = \textbf{Modularity} + \textbf{Validity} + \textbf{Robustness}
\end{equation*}
\begin{description}
    \item[Modularity]
        \FlexCAST achieves modularity by breaking an analysis down into smaller, manageable units, called tasks.
        These tasks are functionals themselves ($F_i$) which can incorporate a dependence on the data $d$ and parameters $p$.
        The entire analysis functional is the composition of the single task functionals ($F= F_n \circ ... \circ F_1$), enabling its alteration through granular analysis steps.
    \item[Validity]
        The validity aspect of \FlexCAST focuses on ensuring that the results produced by a function $f$ remain valid and meaningful when using new data and different parameters in the functional $F$.
        This is achieved through rigorous and automated testing of each individual function $f_i$ and communicating the validity of the results.
        This ensures the trust in and, by extension, publication of a changed analysis.
    \item[Robustness]
        \FlexCAST also aims to increase the robustness of the functional $F$ to maximize the space of validity of $f$.
        Robustness is achieved by ensuring that, within possibly predefined bounds, changes in the data and parameters still yield valid functions, which deliver meaningful physics results.
        This enhances the reliability and applicability of the analysis across various scenarios.
\end{description}

By applying these core principles, \FlexCAST enables several advances in preservation and reinterpretation, such as the optimization of search modes, like adjusting the width of a signal window or retraining a classifier, facilitating updates not only in signal simulations but the entire input data, potentially accommodating improvements in modeling and simulation techniques, and opening up the reinterpretation of analyses for which it was previously impossible, such as data-driven \gls{AD} analyses.

\section{Implementation}
\label{sec:implementation}

This section provides a detailed review of each of the three design principles of {\FlexCAST} - Modularity, Validity, and Robustness - and describes how each principle can be implemented in a scientific data analysis.
While the implementations presented in this paper utilize the Python-based analysis workflow library \gls{law}~\cite{rieger_end--end_2024,rieger_rigalaw_2025}, it is important to emphasize that the core principles of \FlexCAST are agnostic of the workflow and programming language.
Regardless of the specific implementation, it is essential that the collaboration validates, at the latest during the review process of the analysis, that the \FlexCAST principles have been correctly and thoroughly applied to ensure possible future reinterpretations.
Beyond the mentioned principles, each \FlexCAST analysis, including running the analysis and verifying its results, is to be entirely automized, such that not only the analysis code but also its execution are preserved.

\subsection{Modularity}
\FlexCAST follows the principles described in \cite{rieger_end--end_2024} to ensure modularity of the data analysis by transforming the analysis into a collection of functional components which are interconnected within a \gls{DAG}.
These functional components are modular and parametrizable work units, called tasks, which perform a granular operation of the analysis.
Examples of such tasks include a physics event selection, the training of a \gls{ML} classifier, or the statistical inference to quantify the significance of a particular observation in the data.
Each task can have zero, one, or multiple inputs and one or multiple outputs.

We categorize tasks into three distinct types:
Production tasks constitute the core of the analysis, encompassing operations such as data selection, classifier training, classifier application, or statistical inference;
Testing tasks are designed to verify that production tasks behave as intended during the analysis runtime;
And finally, supervision tasks provide additional, non-critical insights into the analysis process, typically through visualization of intermediate results.

The interfaces between tasks within the \gls{DAG} are defined through storage-based targets (e.g. files and directories).
These targets should be unique for a particular set of task parameters.
Many workflow management libraries, such as \law~\cite{rieger_end--end_2024,rieger_rigalaw_2025} and \luigi~\cite{the_luigi_authors_spotifyluigi_2025}, readily allow for a straightforward implementation of these parameter-dependent targets.

Parameters themselves define the behavior of each single task and, by extension, the overall analysis pipeline.
Examples of parameters could be the width of a search window used in bump-hunt algorithms or the number of nodes in the layers of a neural network.
In the case of our implementation using \law and \luigi, parameters are "baked in", meaning they are defined directly within the task code rather than in separate external configuration files.

Parameters can be categorized into two classes: internal and external parameters.
Internal parameters are fixed during the design of the analysis or may be determined dynamically within the analysis itself, such as through optimization algorithms that automatically select the hyperparameters of a neural network.
External parameters remain flexible after analysis design, enabling analysts to explore different analysis parametrizations, such as the used dataset, during the reinterpretation process.

Adopting a \gls{DAG}-based modular computing model provides several additional benefits even in the design stage of the analysis.
Modularity naturally leads to improved isolation between tasks, and by establishing clear abstraction boundaries through encapsulation and modularization, we achieve greater ease of development, reduced entanglement, and fewer errors in the implementation.
Additionally, the \gls{DAG} structure helps to ensure an efficient use of computing resources - when parameters are modified, the \gls{DAG} automatically identifies and reruns only the tasks affected by the changes, thus potentially minimizing the number of computations needed during the reinterpretation process.
Together with the uniqueness requirement of targets, this also paves the way for a future implementation of sharing outputs between different users who want to run the same tasks for their specific reinterpretation needs.
Following \cite{rieger_end--end_2024}, tasks can be run in unique software environments following a “sandboxing” mechanism, which can, for example, run each task in a standalone Docker container~\cite{boettiger_introduction_2015}.
Lastly, modularity is critical for enabling independent and systematic testing of individual tasks, as discussed in the following section.

\begin{figure}
\centering
\tikzset{
    block/.style={rectangle, draw, text width=30mm, text centered,
        rounded corners, minimum height=2.5em, fill=#1},
    block/.default=none,
    line/.style={draw=#1, -Latex},
    line/.default=black,
    every path/.style={line width=1pt}
}
\begin{tikzpicture}[node distance=1cm]
  \node (productionin) at (0,0.25) {Production Input};
  \node[block] (production) at (0,-1) {Production Task};
  \node (productionout) at (0,-3.25) {Production Output};
  \node[block] (test) at (4,-2) {Test Task};
  \node (testout) at (4,-3.25) {Test Output};

  \draw[-latex] (productionin) -| (test);
  \draw[-latex] (productionin) -- (production);
  \draw[-latex] (production) -- (productionout);
  \draw[-latex] (production) |- (test);
  \draw[-latex] (test) -- (testout);
\end{tikzpicture}
\caption{
    This figure illustrates a Production Task and its corresponding Test Task.
    In this example, the Test Task gets the Production Input and the Production Output as inputs.
    The Test Output contains the metrics of the test, including the status.
}
\label{fig:tasks:test}
\end{figure}
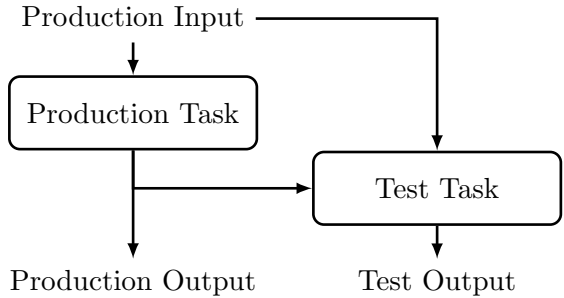

\subsection{Validity}
Because the \FlexCAST functional provides different functions for each set of data and parameters, it is critical to test the validity of the output of each function.
To systematically achieve this, we employ a task-based testing approach according to the task definitions from the previous section.
The test of a task is itself implemented as a task and can be seamlessly integrated into the overall task \gls{DAG}.
This design allows for automatic and modular validation of each production task, ensuring its validity and, by extension, the validity of the entire analysis pipeline.
An important benefit of this approach is that it encourages analysts to explicitly formulate and document their expectations regarding task functionality and methodology.

The testing philosophy adopted in \FlexCAST differs significantly from traditional software testing paradigms.
While conventional software testing typically distinguishes between unit tests (which target small, isolated code units and usually do not rely on external dependencies), integration tests (which evaluate the communication between units), and system tests (which evaluate entire systems or workflows), our task-based testing occupies an intermediate position.
Depending on the complexity and granularity of the tasks, tests range from unit-like tests for simple tasks to more comprehensive, system-like tests for complex, multi-step tasks.

Another distinctive feature of our testing approach is the use of actual runtime parameters and data.
Unlike traditional software tests that often rely on mock data, \FlexCAST tests the behavior of tasks on the actual data and parameters flowing through the analysis pipeline at runtime.
This dynamic testing is particularly crucial for tasks which include the training of \gls{ML} models, as their performance can heavily depend on the specific input data.
As noted in \cite{breck_ml_2017}, testing in \gls{ML} contexts extends beyond code validation and includes testing data schemas, data distributions of model inputs and outputs, and trained models themselves.

\Cref{fig:tasks:test} illustrates a task along with its corresponding test task, detailing the inputs and outputs of each.
Inputs for a test task typically include the outputs of the production task under evaluation, optionally the inputs to that production task, and optionally additional information required for the test.
The outputs of a test are qualitative and quantitative metrics summarizing the task's validity and performance.
We advocate for two central metrics to systematically assess each task:
\begin{description}
    \item[Status]
        This metric provides a qualitative assessment of the task validity, facilitating quick evaluation during the reinterpretation process.
        We propose a simple three-level status indicator (green, yellow, red), indicating a deviation from the expected behavior corresponding to a significance of $\leq1\,\sigma$, $>1$ and $\leq3\,\sigma$ and $>3\,\sigma$.
        Analysts are encouraged to define the status thresholds based on rigorous mathematical expectations (such as underlying probability distributions or calculated p-values), proven assumptions, and best practices (e.g., mathematical approximations that can only be used if certain requirements are fulfilled), or, when necessary, heuristic criteria informed by expert domain knowledge.
    \item[Notes]
        Alongside the qualitative status, we recommend providing detailed quantitative statements or notes that offer a more thorough evaluation of task behavior, helping analysts understand the nature and extent of any deviations or issues encountered if they want to examine a test in more detail.
\end{description}
Explicitly formulating these metrics encourages analysts to transparently document and justify their validation criteria during the analysis design, benefiting both initial and future iterations of the analysis.

\Cref{code:task} shows an exemplary implementation of a task, and \cref{code:test} shows an exemplary implementation of its corresponding test in the Python programming language and using the \law package.
The \texttt{Task} implements a \texttt{parameter} and an \texttt{input}, whereas the \texttt{input} is generated by \texttt{AnotherTask} and the \texttt{answer} is produced via a \texttt{work} function and saved in the \texttt{output}.
The \texttt{Test} infers and saves a \texttt{status} corresponding to the validity of the \texttt{answer} from \texttt{Task}.

\begin{minipage}[t]{.45\textwidth}
\begin{lstlisting}[style=pythoncode,caption=\FlexCAST production task,label=code:task]
class Task(flexcast.ProductionTask):
    parameter = law.Parameter()

    def input(self):
        return SomeInputTask.req(self)

    def output(self):
        return self.target("out")

    def run(self):
        answer = work(
            data=self.input(),
            param=self.parameter,
        )
        self.output().dump(answer)
\end{lstlisting}
\end{minipage}\hfill
\begin{minipage}[t]{.45\textwidth}
\begin{lstlisting}[style=pythoncode, caption=\FlexCAST testing task,label=code:test]
@flexcast.test(Task)
class TestTask(flexcast.TestTask):
    def requires(self):
        return Task.req(self)

    def run(self):
        task_answer = self.input()
        status = test(
            task_answer,
            self.parameter
        )
        self.set_status(status)
        super().save_test()
\end{lstlisting}
\vspace{1cm}
\end{minipage}

Many challenges in \gls{HEP} analyses are addressed through common methods, which are validated by the same criteria.
For example, a standard procedure in many \gls{HEP} analyses involves performing a likelihood fit of a functional model to histogrammed data to determine the final significance of a physical model.
Typically, these fits rely on specific statistical approximations that are valid only under certain conditions.
As an illustrative check, one might verify that each histogram bin contains a sufficient minimum number of events to ensure the validity of these approximations.
Furthermore, physics-specific validations may examine the physical consistency of the data, such as confirming that histograms reflecting systematic uncertainties exhibit two-sided variations around the nominal data.
Taken together, domain-specific tests like these help ensure the validity of an analysis.

Another important testing domain consists of analyses involving \gls{ML} components.
Testing is especially crucial for these analyses, as \gls{ML} tasks frequently connect inputs and outputs across multiple analysis steps, potentially significantly increasing the overall technical debt~\cite{sculley_machine_2014}.
Moreover, testing these tasks is inherently challenging, as their components often rely on non-convex optimization methods that can exhibit non-deterministic behavior due to variations in initialization, data ordering, and stochastic training processes.
Due to the latter, \gls{ML}-based system behavior cannot be precisely specified in advance, and mock testing of outputs is generally infeasible.
\FlexCAST acknowledges this challenge and encourages the user to use a comprehensive testsuite which checks the input data, the training process, and the model itself and explicitly quantifies and incorporates the inherent statistical variability.
Tests can for example include a validation of the inputs to the \gls{ML} training for unexpected artifacts, the monitoring of the evolution of the training loss where its behaviour is predictable, the comparison of the final accuracy on the training and validation datasets, a deliberate overtraining to detect if the model can in principle perform the task, the verification of numerical stability via checking for NaNs and infinities, enforcing bounds on model weights, or ensuring that neuron activations (e.g. non-zero ReLU units) remain withing the expected ranges~\cite{breck_ml_2017}.
Additionally, performance metrics such as the memory usage (VRAM, RAM) and inference times can be monitored to find unexpected anomalies in the training and evaluation process.

To streamline the integration of testing into the analysis, we propose to establish a "check catalogue" that collects best practices and common validation checks depending on the used algorithms.
Such a catalogue can be developed at the level of individual analysis teams or collaboratively across research groups and collaborations, facilitating consistent and rigorous validation standards.
New checks can be incrementally added by analysis teams during the analysis design or together with the publication committees during the analysis review process.

Our implementation of \FlexCAST supports automated generation of comprehensive testing reports based on single tests.
For each test, these reports contain a unique identifier of the testing task, the used parameters, and the resulting metrics consisting of the status, and optional notes or other auxiliary information as shown in \cref{fig:tests:benchmark2}.
By traversing the task \gls{DAG}, the framework automatically identifies and executes all relevant tests.
All metrics are compiled into a final report, which systematically documents the validity of each task and the overall analysis during the reinterpretation process.

Finally, it is important to emphasize that the \FlexCAST testing environment complements, rather than replaces, traditional testing practices.
Analysts are encouraged to continue writing unit tests and test granular functions during the design of the analysis code, as these provide additional, complementary validation beyond the scope of \FlexCAST's task-based testing framework.
Further approaches to software testing in scientific contexts have been covered in~\cite{kanewala_testing_2014}.

\subsection{Robustness}

In the context of \FlexCAST, we define the robustness of a parametrized task as the extent of the parameter space over which this task can successfully operate, meaning it passes its associated tests.
Consequently, a more robust and thus flexible analysis remains valid and functional across a larger region of the overall parameter space of all tasks.
Classical software components, which do not directly depend on varying inputs, tend to be inherently robust by design.
In contrast, tasks involving algorithms that are sensitive to input data - such as fitting procedures or \gls{ML} training - can require explicit measures to ensure robustness.

By leveraging expert domain knowledge and established best practices, several strategies can be employed to enhance robustness, depending on the specific functionality of the production tasks.
Ideally, robustness considerations should directly inform the design of the tasks.

For tasks involving functional fitting procedures, we recommend iterative fitting methods, in which the complexity and flexibility of the fit functions gradually increase~\cite{collaboration_weakly_2025}.
For tasks employing \gls{ML} algorithms, we suggest several practices to enhance robustness, such as augmenting the training data or applying regularization techniques during model training.
Additionally, analysts should aim to select the simplest \gls{ML} models that achieve satisfactory performance, avoiding overly complex and unstable configurations as far as possible.
Another effective robustness strategy specific to \gls{ML} tasks is the use of ensemble methods.
Furthermore, training a base model during the initial analysis phase and subsequently fine-tuning it during reinterpretation can naturally lead to more robust outcomes~\cite{mikuni_solving_2025}.

\section{Case Study}
\label{sec:study}

This section demonstrates the implementation of the \FlexCAST principles for an analysis of LHC-like data.
The chosen analysis strategy is an \gls{AD} analysis, which is particularly well-suited because this type of analysis usually strongly depends on the input data and their reinterpretation is particularly enabled by the new strategies given by \FlexCAST.
\gls{AD} analyses have a growing importance in the investigation of present and future \gls{HEP} data because they fill the gaps between dedicated searches and provide a signal-agnostic approach to discover new and unexpected physics phenomena.

Our case study employs a weakly supervised \gls{AD} technique known as \gls{CWOLA} \cite{metodiev_classification_2017} that searches for a yet unknown particle, whose decay produces a localized peak in a measured invariant mass spectrum.
The \gls{CWOLA} method is a high-dimensional, \gls{ML}-based strategy that can effectively perform an extended bump-hunt in an invariant mass spectrum using many input features.
To perform this search, the measured invariant mass spectrum is partitioned into a \gls{SR}, where a potential signal is hypothesized, and a \gls{SB}, which is assumed to only contain background events.
A classifier is then trained in a weakly supervised manner to distinguish events from the \gls{SR} and \gls{SB}.
Since the main difference between these two regions - apart from the invariant mass - is the potential presence of signal events in the \gls{SR}, the classifier becomes sensitive to the signal features, and applying a threshold on the classifier output can enhance the relative purity of signal events.
Subsequently, a bump-hunt algorithm is applied to the invariant mass spectrum to quantify the significance of the now enhanced excess.
The \gls{SR} and \gls{SB} windows are then systematically slid across the invariant mass spectrum, repeating the classification and bump-hunt for each window definition.
Since the classifier is trained directly on the data and its performance depends on the presence of a signal, this type of analysis is highly sensitive to the input data and data with a different signal or a different signal hypothesis would necessitate an entire rerun of the analysis, in particular retraining the \gls{ML} classifiers.

The \gls{AD} demonstrator in this case study is implemented in a signal-agnostic mode, meaning that the analysis automatically evaluates different datasets and reports their observed significance after the full analysis pipeline.
A user could upload new experimental data or data with an injected signal model to evaluate the analysis sensitivity to their chosen model.

The demonstrator uses the publicly available LHC Olympics (LHCO) R\&D dataset~\cite{kasieczka_lhc_2021,kasieczka_official_2019}, which is commonly used to benchmark \gls{AD} studies in \gls{HEP}.
The dataset contains a background consisting of generic quark-gluon scattering processes that produce dijet events.
As a benchmark, this paper adds a signal process of $W'\rightarrow XY$ with both $X,Y \rightarrow q\bar{q}$, and with the masses $m_{W'}=\SI{3500}{\GeV}$, $m_{X}=\SI{50}{\GeV}$, $m_{Y}=\SI{600}{\GeV}$, which was also used in~\cite{cheng_incorporating_nodate}.
All simulations were performed with \textsc{Pythia} 8.219~\cite{sjostrand_pythia_2006,sjostrand_introduction_2015} and \textsc{Delphes} 3.4.1~\cite{favereau_delphes_2014,mertens_new_2015}.

The reconstructed particles of each event are clustered into jets using the anti-$k_T$ algorithm \cite{cacciari_dispelling_2006,cacciari_fastjet_2012,cacciari_anti-k_t_2008} with a radius parameter of $R=1.0$.
Each event is characterized by the masses of the two jets ($m_{J_1}$, $m_{J_2}$), and the n-subjettiness ratios of the two jets ($\tau_{21}^{J_1}$, $\tau_{21}^{J_2}$, $\tau_{32}^{J_1}$, $\tau_{32}^{J_2}$), whereas the n-subjettiness ratios are defined as $\tau_{ij} = \tau_i / \tau_j$ being particularly sensitive to the two- ($\tau_{21}$) and three- ($\tau_{32}$) prong nature of the jets~\cite{thaler_identifying_2011,thaler_maximizing_2012}.
The two jets from each of the two particles $X$ and $Y$ will mostly be captured each inside a single jet, and their invariant mass \mjj will be close to $m_{W'}$.

Four different benchmark datasets are prepared to systematically evaluate and demonstrate the capability of \FlexCAST to handle different input data scenarios, highlighting its flexibility in analysis reinterpretation.
They are constructed as follows: the \dataset{background} dataset contains approximately 600.000 background events and the datasets \dataset{blackbox 1}, \dataset{2}, and \dataset{3} contain the same amount of background events and the number events of the described signal process corresponding to a significance of 1, 3, and 5\,$\sigma$ in the inclusive \gls{SR}.
Because each dataset is split evenly into 50\% used for training and 50\% reserved for application and inference, reducing the effective sensitivity by a factor of $\sqrt{2}$, the actual injected sensitivity is increased by this factor for demonstration purposes.

The implementation is entirely done in Python and uses \law \cite{rieger_end--end_2024,rieger_rigalaw_2025} to construct and manage the directed acyclic task graph (DAG), and both, the entire production and testing workflows, can be executed via a single command.
To ensure reproducibility, reliability, and ease of deployment, the complete software environment is encapsulated in a single Docker container~\cite{boettiger_introduction_2015}.
This container is automatically built and tested through a Continuous Integration and Continuous Deployment (CI/CD) pipeline hosted on GitLab.
In addition to automating the software build and testing, the GitLab CI/CD pipeline also exemplarily automates the execution of the full analysis workflow, and results and validity reports are automatically generated and presented to the user as pipeline artifacts, which can be readily accessed and reviewed through the GitLab web interface.
The implementation is publicly available via \url{https://gitlab.cern.ch/dnoll/flexcast_demo}.

The production workflow consists of the following tasks:
\begin{description}
    \item[\task{DownloadDataset}] retrieves the input dataset from a remote file source.
    For demonstration purposes, datasets are stored in a Google Drive directory, making them easily accessible from anywhere and eliminating the need for local availability of data before analysis execution.
    The input dataset is defined as an external parameter, allowing for straightforward scanning of multiple datasets.
    \item[\task{DataPreparation}] performs the event selection based on the targeted \mjj range, categorizes events into \gls{SR} and \gls{SB}, and splits the dataset into training and validation subsets.
    \gls{SR} is defined as region of \mjj with a with of \SI{400}{\GeV} as an internal parameter, the \glspl{SB} are defined as regions of \mjj around the \gls{SR} with a width of \SI{400}{\GeV} each.
    \item[\task{ClassifierTraining}] trains the classifiers on the training data to distinguish events from \gls{SR} and \gls{SB}.
    This task has an increased robustness - it not only trains a single classifier but an ensemble of classifiers, whereas the ensemble fluctuates less under the random initialization of a single classifier and the shuffling of the training data.
    The specifications of the used classifier ensemble (used input features, number of classifiers in ensemble, number of nodes in each classifier, number of epochs used in training) are defined as external parameters.
    This task is directly tested by the \task{TestClassifierBias} task as described below.
    \item[\task{ClassifierApplication}] determines a threshold for each trained classifier, and selects events of the data which pass the threshold, resulting in multiple event selections - one per classifier.
    The threshold is determined based on the efficiency of the background rejection, which is defined as an external parameter.
    \item[\task{Histogram}] produces histograms of \mjj for all events that pass the selection of a particular classifier.
    Additionally, the task produces the nominal histogram, which is used in the downstream production tasks, as the mean of the individual histograms from all classifiers in the ensemble.
    All histograms in the demonstrator contain nine bins, three in each of the \glspl{SB} and three in the \gls{SR}.
    This task is directly tested by the \task{TestHistogramSpread} task and the \task{TestHistogramStats} as described below.
    \item[\task{Fit}] performs a fit of an exponential function to the counts of the \mjj histogram in the \gls{SB}.
    The task enhances robustness by iteratively fitting different functions, progressively increasing their complexity and flexibility, and stopping when a fit function with a chi-squared value divided by the number of degrees of freedom below 1 is achieved or the most complex function has been reached.
    The statistical uncertainty from the data counts in each bin and the spread from the different classifiers drive the uncertainty of the fit.
    This task is tested by the \task{TestFit} task, which is described below.
    \item[\task{Inference}] extrapolates the fitted background distribution from the \gls{SB} into the \gls{SR} and calculates the local statistical significance of any observed excess of the inclusive number of data events in the \gls{SR} to the inclusive number of estimated background events in the \gls{SR} considering the statistical and fit uncertainties.
    \item[\task{InferenceScan}] executes the complete workflow described above for multiple sliding window definitions of \gls{SR}, systematically scanning the \mjj spectrum.
\end{description}

The validity of production tasks is demonstratively ensured by a set of testing tasks.
While these tests illustrate the capabilities and principles of \FlexCAST, we do not claim that every component of the demonstration analysis is exhaustively tested.
The testing workflow consists of the following tasks:
\begin{description}
    \item[\task{TestClassifierBias}] tests the output of the \task{ClassifierTraining} task by checking the bias of the response of the classifier ensemble.
    The task compares the cross-validation loss of the classification on the training and the validation dataset.
    A relative difference between training and validation loss of less than 1\% indicates a green status, a relative difference between 1\% and 10\% indicates a yellow status, and for everything above 10\%, the test is considered failed and the status is red.
    The values are a simple heuristic employed for the demonstrator analysis.
    \item[\task{TestHistogramSpread}] tests the output of the \task{Histogram} task and implicitly verifies the spread in the classifier ensemble from \task{ClassifierTraining}.
    The task verifies that the histograms, which are based on the different classifier selections, are consistent within the ensemble of classifiers.
    For every classifier $i$ and every bin $j$, the task calculates the z-score $z_{ij}$ with the counts from the classifier ($p_{ij}$) and the mean ($\mu_{ij}$) and standard deviation ($\sigma_{ij}$) of the counts from all other classifiers as $z_{ij} = (p_{ij}-\mu_{ij})/\sigma_{ij}$.
    The sum over the z-scores of all classifiers and bins $z = \sum_i\sum_j z_{ij}$ is used to calculate a final p-value of the test.
    If the p-value corresponds to a significance of $\leq1\,\sigma$ the status of the test is green, if it corresponds to a significance of $>1\,\sigma$ and $\leq3\,\sigma$, the status of the test is yellow, if it corresponds to a significance $>3\,\sigma$, the test is considered failed and its status is red.
    \item[\task{TestHistogramStats}] tests the output of the \task{Histogram} task and ensures that every bin in the histogram is filled with a minimum number of events.
    If every bin contains ten or more events, the status of the test is green; if not, the test is considered failed, and the status is red.
    \item[\task{TestFit}] evaluates the quality of the fit of the exponential function from the \task{Fit} task.
    The task calculates the $\chi^2$ between the fitted function and the observed count in the \gls{SB}, considering the statistical and classifier uncertainty of the observed counts and the uncertainties of the fitted function.
    This $\chi^2$ value and the number of degrees of freedom from the fit are used to calculate a p-value.
    If the p-value corresponds to a significance of $\leq1\,\sigma$ the status of the test is green, if it corresponds to a significance of $>1\,\sigma$ and $\leq3\,\sigma$, the status of the test is yellow, if it corresponds to a significance $>3\,\sigma$, the test is considered failed and its status is red.
\end{description}

In addition to production and testing tasks, the implementation of the demonstrator includes supervision tasks that generate visualizations of intermediate results.
As they are not necessary for a scientifically rigorous result, they are not discussed in more detail.

\begin{figure}
\centering
\begin{subfigure}{0.5\textwidth}
    \centering
    \includegraphics[width=\linewidth]{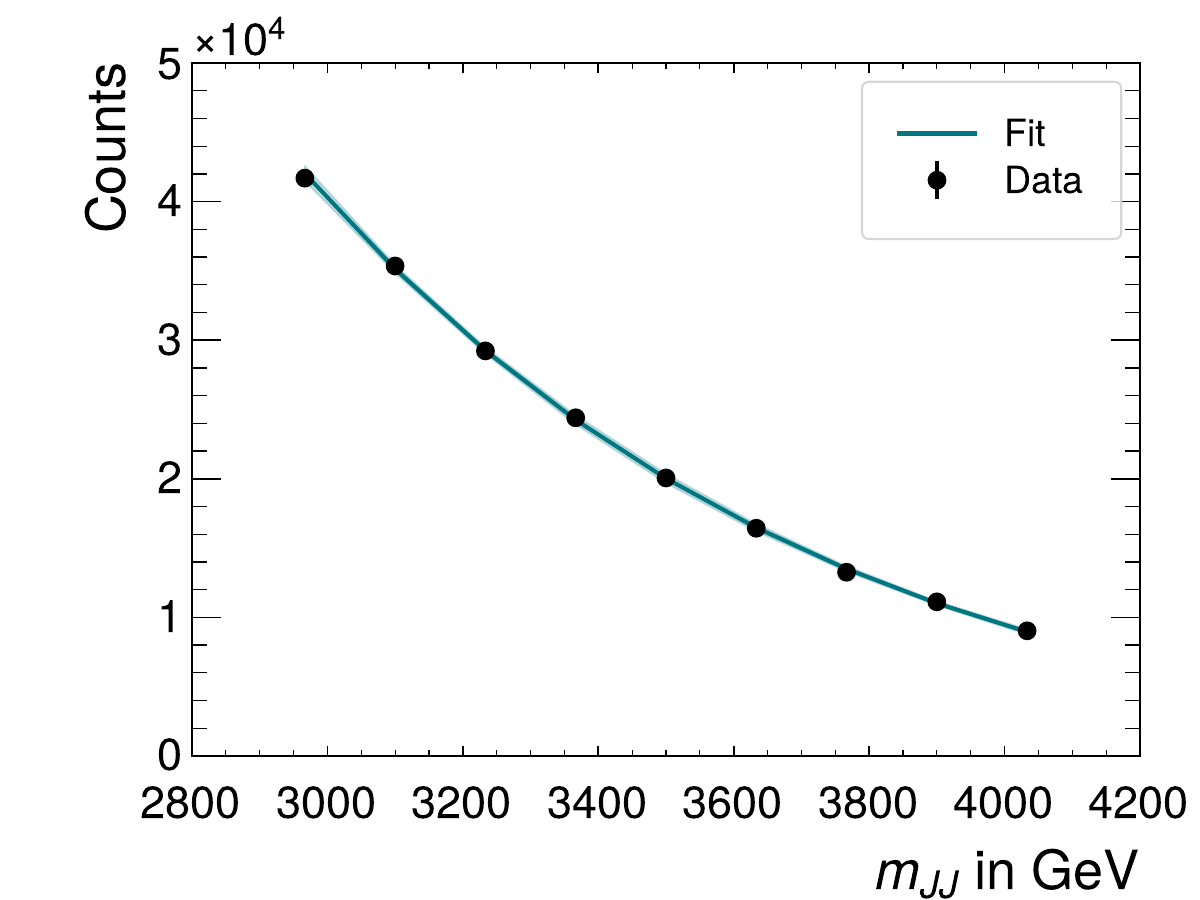}
    \caption{No cut}
    \label{fig:mjj:0p001}
\end{subfigure}%
\begin{subfigure}{0.5\textwidth}
    \centering
    \includegraphics[width=\linewidth]{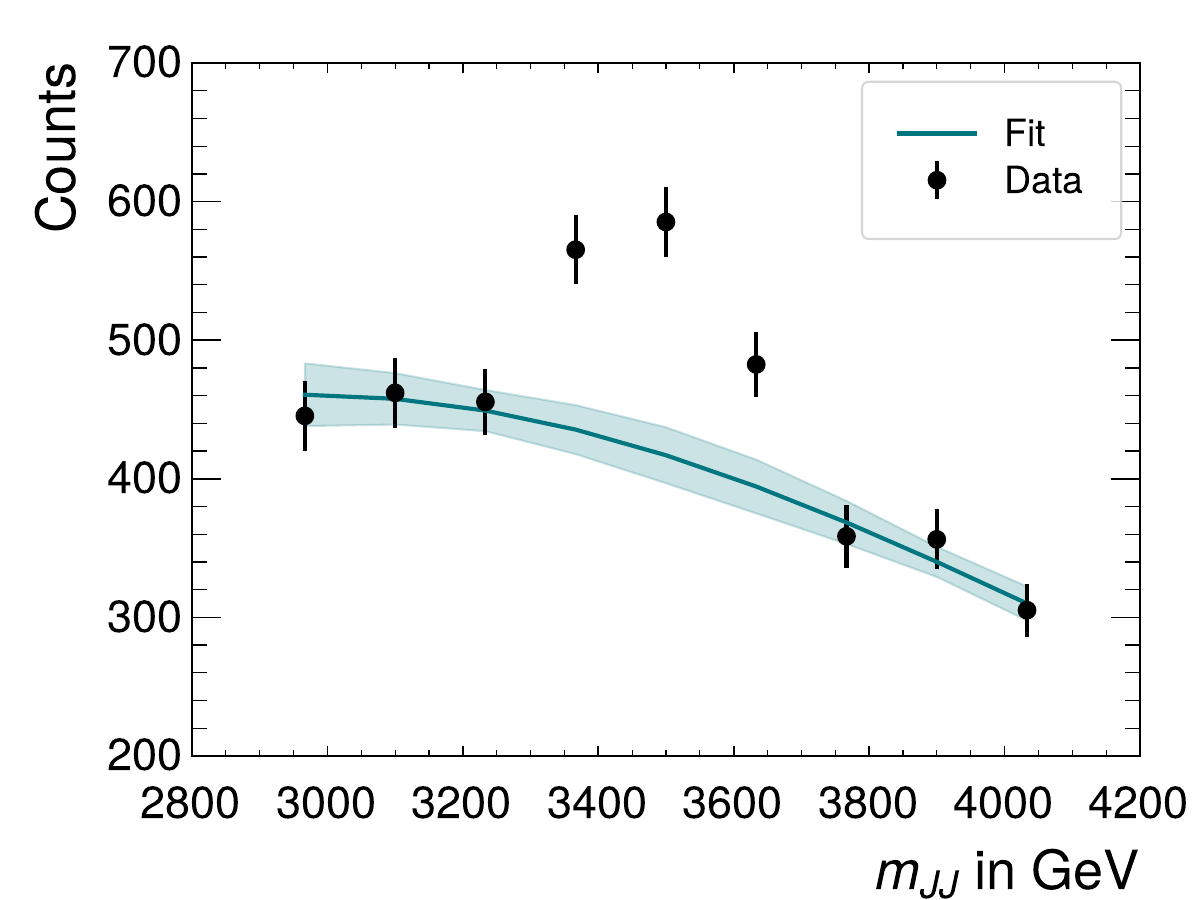}
    \caption{98\% cut}
    \label{fig:mjj:0p98}
\end{subfigure}
\caption{
    Spectrum of invariant mass before and after the classifier selection.
    The figures show the results from \gls{SR} $[3300,3700]\,\si{\GeV}$, which contains the \gls{BSM} signal.
    The used dataset is benchmark dataset 2 with an initial injected signal significance of 3\,$\sigma$ in the \gls{SR}.
}
\label{fig:mjj}
\end{figure}

\Cref{fig:mjj:0p001} shows the initial distribution of the invariant mass of the two jets \mjj in the dataset \dataset{blackbox 2}.
\Cref{fig:mjj:0p98} shows the distribution of the same feature, after the training of the classifier in the \gls{SR} $[3300,3700]\,\si{\GeV}$ and applying it on the entire range with a background rejection of 98\%.
Both figures include a background estimation using an exponential function with four parameters, which is fitted to the event counts in the \gls{SB}.
It is visually evident as a peak in \cref{fig:mjj:0p98} that the \gls{AD} strategy increases the purity of the signal.

\begin{figure}[htbp]
    \centering
    \includegraphics[width=0.5\linewidth]{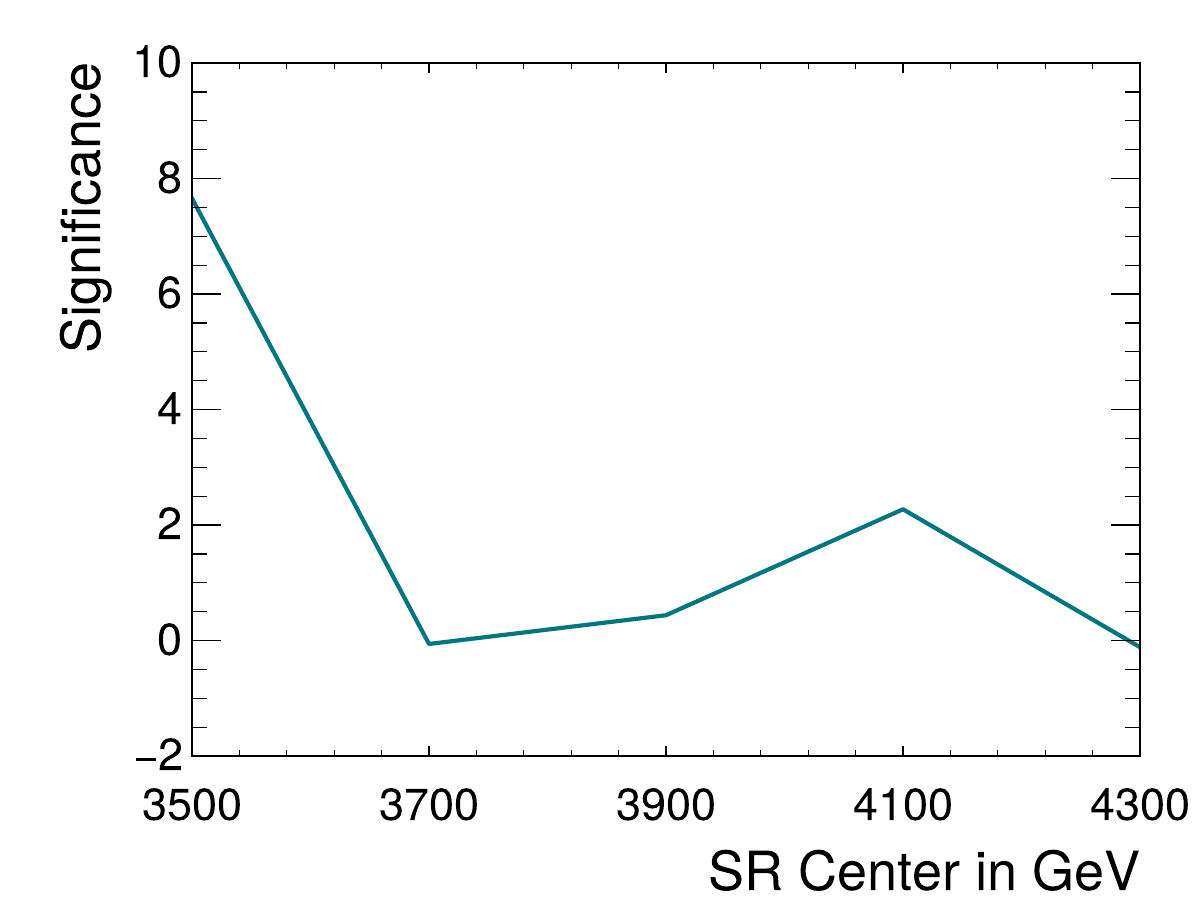}
    \caption{
        Final significance of the analysis pipeline for the different signal regions.
        The used dataset is benchmark dataset 2 with an initial injected signal significance of 3\,$\sigma$ in the \gls{SR}.
        This initial significance is improved to almost 8\,$\sigma$, indicating a successful working of the analysis.
    }
    \label{fig:significances}
\end{figure}

\Cref{fig:significances} shows the measured local significances for each scanned \gls{SR} in the \dataset{blackbox 2} dataset.
The analysis methodology enhances the significance of the injected signal in the corresponding \gls{SR} from the original 3\,$\sigma$ to almost 8\,$\sigma$, while simultaneously avoiding significant false-positive detections in any other \glspl{SR}.
A similar behavior is observed in the \dataset{blackbox 3} dataset.
Neither the background-only dataset nor the \dataset{blackbox 1} dataset shows significant excesses.
The background-only dataset contains no signal by design, while the signal injected in \dataset{blackbox 1} is too small to be detected by the employed \gls{AD} strategy, an effect that has been studied previously~\cite{hallin_classifying_2022}.

\begin{figure}
\centering
\begin{subfigure}{0.398\textwidth}
    \centering
    \includegraphics[width=0.9\linewidth]{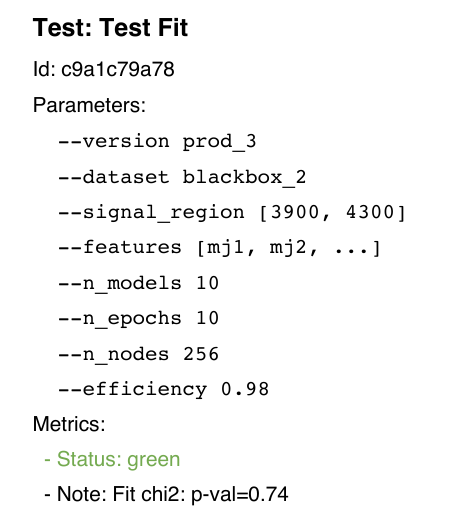}
    \caption{}
    \label{fig:tests:benchmark2}
\end{subfigure}%
\begin{subfigure}{0.601\textwidth}
    \centering
    \includegraphics[width=0.9\linewidth]{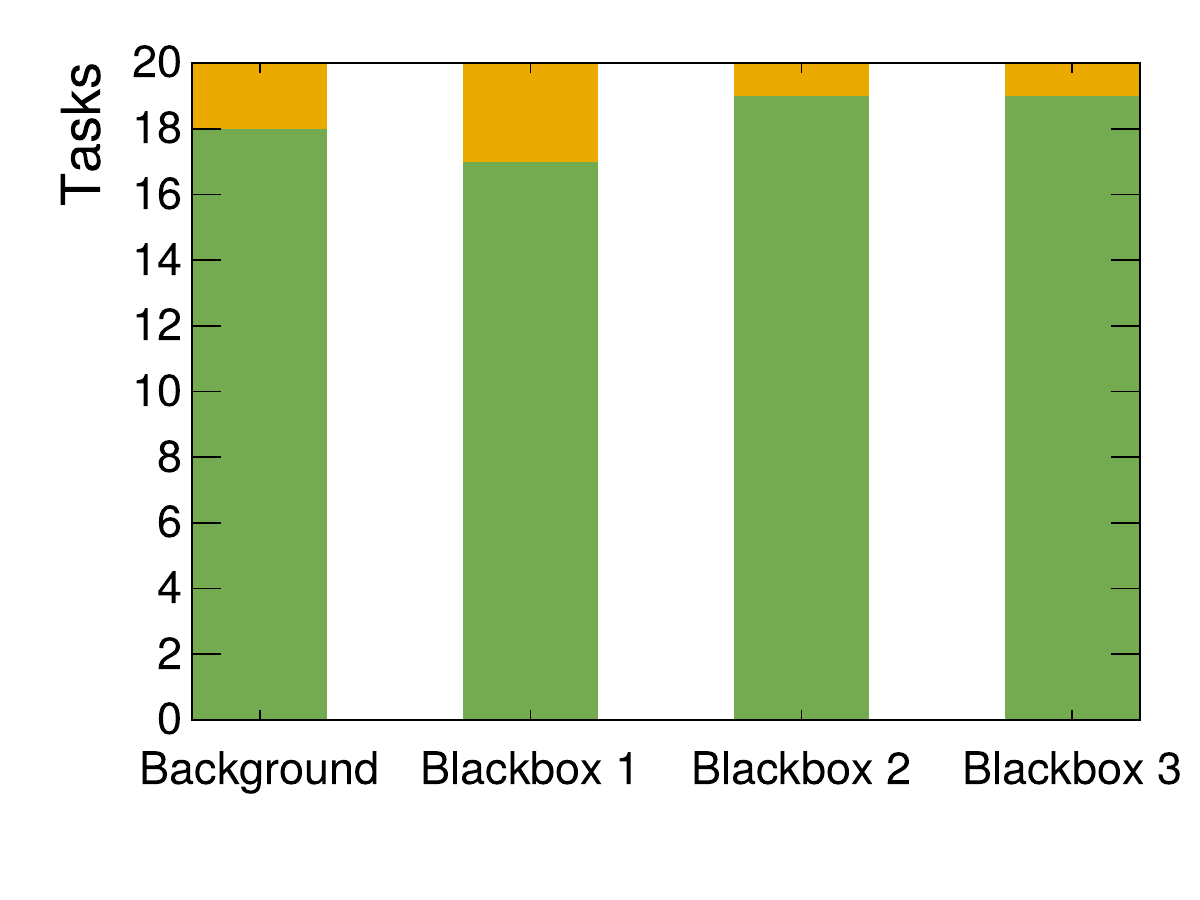}
    \caption{}
    \label{fig:tests:all}
\end{subfigure}
\caption{
    Outputs of testing tasks in the \FlexCAST framework.
    (a) shows the output of \task{TestFit} run on the \dataset{benchmark 2} dataset, stating the used parameters and resulting metrics.
    The test finishes with a green status with a p-value of 0.74.
    (b) Summary of the test statuses from running all tests on all benchmark datasets.
    Most of the tests finish with a green status, some tests finish with a yellow status, which is expected and explained in the main text.
    No failed tests (red status) were observed, indicating the robustness of the analysis tasks.
}
\label{fig:tests}
\end{figure}

While the last paragraphs showed the flexibility of the analysis in its application to different input datasets and parametrizations, the following paragraph will showcase the testing workflow of the demonstrator analysis.
\Cref{fig:tests:benchmark2} shows an exemplary output of \FlexCAST for the testing task \task{TestFit} executed on \dataset{blackbox 2} .
All parameters are included in the output for a straightforward reference to the task and to give the ability to quickly check its output or rerun it.
The output gives a comprehensive overview of the executed task and its validity, given by the status and the notes.
\Cref{fig:tests:all} shows the testing status of all analysis tasks after running the analysis on all benchmark datasets.
Most of the tests are passed with a green status.
Given that a yellow status corresponds to a deviation from the expected behavior with a significance of $>1\,\sigma$ and $\leq3\,\sigma$, we expect approximately 30\% of the tasks to have a yellow status.
However, because some tasks in this demonstrator, such as \task{TestHistogramStats}, employ simpler heuristical criteria for the tests which do not directly translate to a strict statistical measure, the number in \cref{fig:tests:all} is lower.

While the case study incorporates many key elements of a state-of-the-art \gls{HEP} analysis - including data preprocessing, \gls{ML} techniques, and statistical inference - we acknowledge that the computational tasks presented here are relatively lightweight and that the tests employed in the demonstrator are not necessarily exhaustive.
Despite its methodological simplicity, to the best of the authors' knowledge, this \FlexCAST demonstrator represents the first fully reinterpretable implementation of an \gls{AD} analysis on LHC-like data.

\section{Conclusion and Outlook}
\label{sec:conclusion}

In this paper, we introduced \FlexCAST, a novel framework for flexible scientific data analysis that enables comprehensive changes to both the input data and the analysis parametrization during the reinterpretation process.
This flexibility is realized through three central design principles: Modularity, Validity, and Robustness.
Modularity provides a structured and flexible implementation of scientific data analyses, enabling automatic execution and preservation of the design of the analysis and its execution workflow.
Validity ensures that new analysis configurations are automatically validated, facilitating the reliable extraction of new results.
Robustness enhances the reliability of analyses to extend the validly covered data and parameter space.
We demonstrated the \FlexCAST principles through a concrete implementation of a state-of-the-art \gls{AD} analysis on LHC-like data.

While \FlexCAST offers significant advantages, we acknowledge several challenges and areas for future development.
Compared to \RECAST, \FlexCAST inherently demands greater computational resources, as it involves rerunning larger parts of an analysis rather than minimal reprocessing limited to a changed signal model.
This increased resource demand is an intrinsic consequence of the flexibility provided, particularly when analyses depend strongly on large parts of the input data.
Future strategies, such as fine-tuning of existing \gls{ML} models developed during the original analysis, may help mitigate this resource demand, but these approaches require careful consideration and implementation.

A logical next step for \FlexCAST is the integration into established infrastructures for analysis preservation and execution, such as \REANA~\cite{simko_reana_2019}.
Additionally, the reinterpretation workflow could greatly benefit from a closer integration of software and hardware resources, including but not limited to access to cluster computing (e.g., via Slurm or HTCondor), hardware acceleration, persistent and shared data targets, and user-friendly graphical interfaces.
Although some of these features are already partially integrated into \FlexCAST, their continued development and effective deployment represent exciting technological challenges for future work.

In conclusion, \FlexCAST provides a powerful and flexible framework for scientific data analysis, enabling comprehensive reinterpretation through changes in input data and analysis parametrization.
Given its versatility, the framework has the potential to significantly impact a broad range of scientific analyses, opening new possibilities for efficient and reliable analysis reuse and reinterpretation across scientific domains.

\acknowledgments
We gratefully acknowledge the authors and contributors of existing reinterpretation approaches in \gls{HEP}, particularly the \RECAST and \law teams, whose foundational work has significantly advanced analysis preservation practices in our field and enabled the development of \FlexCAST.
This work is supported by the U.S. Department of Energy (DOE), Office of Science under the contracts No. DE-AC02-05CH11231 and No. DE-AC02-76SF00515.
This research used resources of the National Energy Research Scientific Computing Center, a DOE Office of Science User Facility supported by the Office of Science of the U.S. Department of Energy under Contract No. DE-AC02-05CH11231.

\printbibliography

\end{document}